\documentstyle[12pt,titlepage]{article}
\newcommand{\ba}{\begin{array}}
\newcommand{\ea}{\end{array}}
\newcommand{\bd}{\begin{displaymath}}
\newcommand{\ed}{\end{displaymath}}
\newcommand{\be}{\begin{equation}}
\newcommand{\ee}{\end{equation}}
\newcommand{\bea}{\begin{eqnarray}}
\newcommand{\eea}{\end{eqnarray}}

\def\bra{\langle}
\def\ket{\rangle}

\def\g{\gamma}

\def\m{\mu}

\def\G{\Gamma}

\def\p{\pi}

\begin{document}
\begin{titlepage}
\vspace*{0.5truein}

\begin{flushright}
\begin{tabular}{l}
MRI-PHY/21/94 \\
November, 1994
\end{tabular}
\end{flushright}
\vskip .6cm

\begin{center}
{\Large\bf LIGHT NEUTRALINOS IN B-DECAYS \\[0.5truein]}
{\large Rathin Adhikari$^1$ and Biswarup Mukhopadhyaya$^2$\\}
Mehta Research Institute \\
10 Kasturba Gandhi Marg\\
Allahabad - 211 002, INDIA \\
\end{center}
\vskip .5cm

\begin{center}
{\bf ABSTRACT}\\
\end{center}

We consider the decays of a $B_s$-meson into a pair of
lightest supersymmetric particles (LSP) in the minimal supersymmetric
standard model. It is found that  the parameter space for light LSP's
in the range of 1 GeV can be appreciably constrained by looking for
such decays.

\hspace*{\fill}
\vskip .5in

\noindent
$^{1}$E-mail : ~~rathin@mri.ernet.in  \\
$^{2}$E-mail :biswarup@mri.ernet.in

\end{titlepage}

\textheight=8.9in

The study of B-mesons is hoped to be a rather fruitful adventure in the
years to come \cite{Rev1,Rev2}. B-factories working at the $\Upsilon
(4s)$ -resonance
can produce upto $10^{7-8}$ $B \overline{B}$ -pairs which are mixtures of
$B^{0}_{d} \overline{B^{0}_d}$ and $B^{+} B^{-}$. In addition, if the beams
are tuned just above the $B_s$ -threshold, then $B^{0}_{s}
\overline{B^{0}_s}$ -
pairs can be also present copiously among the products. In addition to
giving insights into the hadronization of heavy quarks, the various things
that B-factories can probe include CP-violation in B-decays, precise
determination of the quark mixing matrix and indirect evidence of physics
beyond the standard model through loop-induced B-decays.

In this note we want to point out that it is also possible to explore
{\it direct} signals of non-standard physics through B-decay experiments
at least in a certain area of the parameter space. In this context we
focus our attention on the supersymmetric (SUSY) extension of the standard
model \cite{susy}. As we all know, lower limits on the masses of
superparticles already
exist in the literature.  The limits on the squark
and the gluino masses as inferred from experiments at the Fermilab Tevatron
are especially stringent, ranging upto the 150-200 GeV range \cite{sqlim}.
However,
beacuse of the necessity to eliminate backgrounds, events with very soft
final states as well as those with little missing transverse energy $\not{E}$
have to be left out. Thus a light gluino ($\approx 5 \; $ GeV or less) can
still
escape detection in such experiments. In such a case, the squarks can
directly decay into gluinos whose decay products may be degraded enough to
be lost among the backgrounds, thereby relaxing the squark mass limits as well.
Thus a window \cite{window}, although controversial \cite{contro}, still
exists with the gluino
in the 2.5-5 GeV range and the squarks with masses around 70 GeV or above.
If the gluino is so light, then according to most viable models the lightest
supersymmetric particle (LSP) will have to be even lighter. Various efforts
to close this window in direct or indirect ways have gone on in recent
times \cite{close}.
Side by side, a light gluino has been claimed  to be instrumental in causing
better agreement between theory and experiment in the evolution of the
strong coupling $\alpha_s$ \cite{lgluino}. Together with other attempts to
theoretically
justify such a scenario (for example, by postulating radiative gaugino
masses \cite{rgaugino}),
the option of light sparticles still remains a matter of lively interest.

Here we address the following question: in a light sparticle scenario,
can a neutral B-meson decay invisibly into a pair of LSP's ?
If that indeed be the case, then, provided that a substantial number of such
decays in a B-factory is pedicted, it will be possible to constrain the
SUSY parameter space  from the viewpoint of light LSP's. Since in most models
the light LSP is the lightest neutralino, we also limit ourselves to that
choice here. Furthermore, such a light LSP is predominantly a photino
state, as can be seen, for example, by taking recourse to a SUSY theory
motivated by Grand Unified Theories (GUT) \cite{GUT}. In such a case the range
in the parameter space that is allowed by LEP experiments and is simultaneously
compatible with a light gluino corresponds to $\mu \approx -50 \;$ to $
- 100 \; $GeV
and $\tan \beta \approx 1.0-1.8$, $\mu$ and $\tan \beta$ being respectively
the Higgsino mass parameter and the ratio of the scalar vacuum expectation
values. On diagonalisation of the neutralino mass matrix containing
parameters in the above range, the LSP turns out to be almost entirely
the photino state.

The process of our concern is the decay
$B_{s}\longrightarrow  \chi^0_1 \chi^0_1$
where $\chi^0_1$ is the LSP. Such an invisible final state state has no
standard model backgound, since the only candidates for an
invisible final state can be the neutrinos whose near-masslessness suppresses
the decay from helicity considerations. At the quark level, the SUSY process
corresponds to $b\longrightarrow s \chi^0_1 \chi^0_1$. Interestingly, such
a flavour-changing neutral current (FCNC) process can be allowed at the
tree-level \cite{FCNC}. This is because the left squark mass matrices are not
simultaneously diagonal with the quark mass matrices. For example, in a
basis where the  charge-1/3 quark mass matrix is diagonal, the charge -1/3
left squark mass matrix is given by

\bea
M_{L_{\tilde{d}}}^2 =
\left( m_L^2 \, {\bf{1}} + m_{\hat{d}}^2+c_0 \, K \, m_{\hat{u}}^2 \,
K^{\dag}
\right)
\eea

\noindent
where $m_{\hat{d}}, m_{\hat{u}}$ are the diagonal down-and up-quark mass
matrix respectively, and K is the Kobayashi-Maskawa matrix. $m_L$ is a
flavour-blind SUSY breaking parameter that sets the scale
of squark masses.  We neglect here left-right mixing
among squarks which can potentially contribute to the off-diagonal blocks.
The term proportional to $m_{\hat{u}}^2$  arises
out of quantum corrections to the left-squark masses induced by up-type Yukawa
couplings. In a scenario where the SUSY is embedded in a higher structure
\cite{hsusy},
the evolution of the soft SUSY-breaking terms from the higher scale to the
scale of the electroweak symmetry makes such quantum corrections
particularly important. Consequently, $M_{L_{\tilde{d}}}^2$ is not
diagonal in this basis, thereby entailing the occurence of flavour
violation in squark-quark-neutralino (or gluino) interactions.
Tree-level graphs (fig. 1) contributing to
$B\longrightarrow  \chi^0_1 \chi^0_1$ originate from this kind of an
interaction. The corresponding term in the lagrangian is

\bea
{\cal L}_{q\tilde{q}\chi_i^0}  =
-{\sqrt{2}} \, g \, \sum_{ij} \, {\tilde q}_j \, {\bar{\chi}_i^0} \, \left[
\, \tan{\theta_w} \, e_j N_{i2}^\star \,
\G_{jk} \,  {{1-\g_5} \over 2}
\right] q_k \, + \, h.c.
\eea

\noindent
where $\G_{jk}$ is the (jk)-th element of the unitary matrix that
diagonalises $M_{\tilde{d}}^2$ in equation(1).
N is the neutralino mixing matrix.

For $b \longrightarrow s$, the element $\G_{23}$ is important. Its
value depends on $m_t$ and $c_0$. In view of the recent results from the
Fermilab Tevatron, we have chosen $m_{t} \; = \; 170 \;$ GeV here. The value
of $c_0$ is model independent; however, as recent estimates indicate, a
value around 0.01 or slghtly above is likely even from a rather conservative
point of view \cite{c0}. Here we write $\G_{23} = c K_{23}$, where c is a
function
of $c_0$. Explicit diagonalisation of the squark mass matrix reveals that
c lies in the range $\approx$ 0.5-0.9 for $c_0  \approx  0.01-0.001$.
Thus the mixing elements relevant for our puopose are quite close in
magnitude to those of the Kobayashi-Maskawa matrix.

Using equation (2) in the limit neuralino $\approx$ photino, the
Fierz-transformed quark level matrix element for
$b(p_0)\longrightarrow s(p_3) \; \chi^0_{1} (p_2) \; \chi^0_{1} (p_1) $ is
given by

\bea
{\cal M} = \; - {e^2 \; c \; V_{23} \over 18 \; {m_{\tilde{q}}}^2}\;
\left [
\; {\bar s}(p_3)\; \g^\m \;(1-\g_5)\;b(p_0)\;{\bar u}(p_1)
\;\g_\m \;(1+\g_5)\; v(p_2) \; - \right. \nonumber \\  \left.
\;{\bar
s}(p_3)\;\g^\m \;
(1-\g_5)\;b(p_0)\; {\bar u}(p_2)\;\g_\m \;(1+\g_5)\; v(p_1)
\; \right ]
\eea

\noindent
where $m_{\tilde{q}}$ is the common squark mass.

Next, one has to use

\bea
\bra 0|{\bar s}(\; \g^\m \;(1-\g_5)\;b)|B_{s}^0 \ket \!=\! f_{B_s} q_{\mu}
\eea

\noindent
where $q $ is the four-momentum of the decaying $B_s$, and
$f_{B_s}$ is the $B_s$-decay constant.

The two-body decay-width is given by

\bea
\G \; = {g^4 \; \sin^4 \theta_w \; {|  V_{23}  | }^2 \; \left(  c^2
f^2_{B_s}  \right)  \over 216 \; \p \; {m_{\tilde q}}^4 } \; m^2 \;
{\left( \; {m_B}^2 \; - 4\; m^2 \right)}^{1/2}
\eea

\noindent
$m$ being the mass of the LSP.

In figure 2 we show the dependence of the branching ratio for
$B\longrightarrow  \chi^0_1 \; \chi^0_1$ on the LSP mass. Here we have taken
$m_{\tilde{q}}$ = 80 GeV, which is within the allowed region of the
parameter space in this scenario. The branching ratio is presented in
units of $c^2 \; f^{2}_{B_s}$.

If one adheres to a scenario inspired by GUT's, then, provided that the light
gluino window is in the 2.5-5 GeV range, one might expect the corresponding
window for the light LSP in the range  0.4-1 GeV. If one looks at the graph,
then this range corresponds to a branching ratio of
$(10^{-4} - 10^{-5}) \; c^2 \; f^{2}_{B_s} \; {\mbox GeV}^{-2}$.
As we have already remarked, with
a heavy top quark the value of c turns out to be quite close to 1. As for the
parameter $f_{B_s}$, its value, although not completely known yet, can
be expected to lie in the range 0.2-0.3 GeV \cite{fBs}. Depending on this, a
branching
ratio of O($10^{-5} - 10^{-7}$) can be expected for the invisible
channel. If an accumulaion of $10^8$ events takes place in a B-factory,
then the number of such decays could be sizable. Moreover, if one
wants to free oneself from the shackles of GUT's and restrict
light LSP's from a phenomenological point of view, then it is possible
to put limits in the range of 1-2 GeV as well, since the branching ratio
is even higher in that range.

It may be noted here that $B_s$-mesons and not $B_d$'s are going to be
useful for the above purpose. This is because for the latter to decay
invisibly, contributions have to come at the quark level from
$b\longrightarrow d \;  \chi^0_1 \; \chi^0_1$. This decay width would thus be
suppressed compared to the $B_s$-decay case by a  factor $(K_{td}/K_{ts})^2$.

The experimental problems that one can foresee are perhaps the difficulty of
obtaining an accurate estimate of the proportion of the $B_s$-mesons in the
admixture of $B^{\pm}$ and $B^0_d$. For the kinds of decay discussed here,
one has to identify one $B_s$ through decay channels such as $ \; D_s \;\pi$
or $ \; J / \psi \; \phi$ \cite{mixBs}, and look for those cases when the
other one in the
pair becomes invisible. There again, the question to address is that of
separating those events where tagging of one $B_s$ is simply unsuccessful.

In summary, we have considered the contributions to the decay width of a
$B_s$-meson from a pair of LSP's, which can make the $B_s$ invisible. The
estimate is model-independent, apart from the introduction of a
phenomenological parameter to quantify the extent of neutral flavour
violation. The prediction shows the feasibility of imposing independent
constraints on the parameter space of light LSP's. Looking for such
invisible decays may thus be an interesting challenge in B-factory
experiments.

\newpage
\centerline {\large {\bf Figure Captions}}

\hspace*{\fill}

\hspace*{\fill}

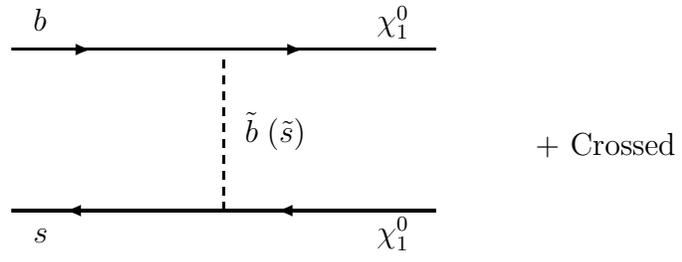
Figure 1:

\noindent
The tree-level contributions to $b\longrightarrow s \chi^0_1 \chi^0_1$.
In addition there will be crossed diagrams where the four-momenta of
the LSP's are interchanged.

\vskip .25in

Figure 2:

\noindent
The branching ratio for invisible $B_s$-decay (in units of
$c^2 \; f^{2}_{B_s}$) plotted against the LSP mass.
$m_{\tilde{q}}\;=\;80$ GeV.


\newpage

\hspace*{\fill}

\hspace*{\fill}

\hspace*{\fill}


\begin{picture}(150,250)(-100,-160)
\thicklines
\put(22,4){\line(1,0){20}}
\put(52,4){\vector(-1,0){10}}
\put(52,4){\line(1,0){70}}
\put(132,4){\vector(-1,0){10}}
\put(132,4){\line(1,0){50}}
\put(22,65){\vector(1,0){30}}
\put(52,65){\line(1,0){50}}
\put(102,65){\vector(1,0){30}}
\put(132,65){\line(1,0){50}}
\put(102,4){\line(0,1){3}}
\put(102,10){\line(0,1){3}}
\put(102,16){\line(0,1){3}}
\put(102,22){\line(0,1){3}}
\put(102,28){\line(0,1){3}}
\put(102,34){\line(0,1){3}}
\put(102,40){\line(0,1){3}}
\put(102,46){\line(0,1){3}}
\put(102,52){\line(0,1){3}}
\put(102,58){\line(0,1){3}}
\put(160,-8){\mbox {$\chi^0_1$}}
\put(160,72){\mbox {$\chi^0_1$}}
\put(30,-8){\mbox {$s$}}
\put(30,72){\mbox {$b$}}
\put(110,30){\mbox {${\tilde b \; ({\tilde s})} $}}
\put(220,25){\mbox {+ Crossed}}
\put(100,-70){\mbox {\Large FIG. 1}}
\end{picture}

\end{document}